\documentclass[final,3p,times,twocolumn]{elsarticle}

\usepackage{amssymb}
\usepackage{amsmath}

\usepackage{array}
\usepackage{tabularx}
\usepackage{booktabs}
\usepackage{hyperref}
\usepackage{xurl}
\usepackage{xcolor}

\journal{Future Generation Computer Systems}

\begin{document}

\begin{frontmatter}



\title{Designing and Implementing a Comprehensive Research Software Engineer Career Ladder: A Case Study from Princeton University}


\author{Ian A. Cosden\corref{cor1}}
\ead{icosden@princeton.edu}
\cortext[cor1]{Corresponding Author}
\author{Elizabeth Holtz}
\ead{eparham@princeton.edu}
\author{Joel U. Bretheim}
\ead{jbretheim@princeton.edu}

\affiliation{organization={Research Computing, Princeton University},
            city={Princeton},
            state={NJ},
            postcode={08544},
            country={USA}}

\begin{abstract}
Research Software Engineers (RSEs) have become indispensable to computational research and scholarship.
The fast rise of RSEs in higher education and the trend of universities to be slow creating or adopting models for new technology roles means a lack of structured career pathways that recognize technical mastery, scholarly impact, and leadership growth.
In response to an immense demand for RSEs at Princeton University, and dedicated funding to grow the RSE group at least two-fold, Princeton was forced to strategize how to cohesively define job descriptions to match the rapid hiring of RSE positions but with enough flexibility to recognize the unique nature of each individual position.
This case study describes our design and implementation of a comprehensive RSE career ladder spanning Associate through Principal levels, with parallel team-lead and managerial tracks.
We outline the guiding principles, competency framework, Human Resources (HR) alignment, and implementation process, including engagement with external consultants and mapping to a standard job leveling framework utilizing market benchmarks.
We share early lessons learned and outcomes including improved hiring efficiency, clearer promotion pathways, and positive reception among staff.
\end{abstract}




\end{frontmatter}



\section{Introduction}
 \label{introduction}
Modern research increasingly depends on robust, performant, and sustainable software \cite{us-software}. 
As evident by their growing numbers in research environments \cite{career-entry-points}, Research Software Engineers are increasingly being recruited to provide embedded, long-term collaborations that bring software engineering rigor, performance optimization, and domain-informed algorithm design to research teams.
Princeton’s central RSE group, housed within the Research Computing department in the Office of the Dean for Research (ODFR), complements traditional academic research groups by embedding RSEs across departments to deliver long-term software development, coding standards, domain expertise, and performance tuning \cite{princeton-rse-group}. 
The goal is explicit: to produce efficient, scalable, and sustainable research code that accelerates scientific and scholarly discovery.
Despite their strategic importance, RSE roles in academia often suffer from unclear job structures, inconsistent or non-existent promotion criteria, and difficulty aligning either scholarly contributions or software outputs with traditional HR frameworks \cite{us-rse-adsa-career-guidebook}.
Princeton’s initiative sought to address these gaps by creating a transparent, competency-based career ladder aligned to institutional HR structures and market benchmarks, while preserving the academic character and research nature of RSE work.

\begin{figure*}[t]
    \centering
    \includegraphics[width=0.8\textwidth]{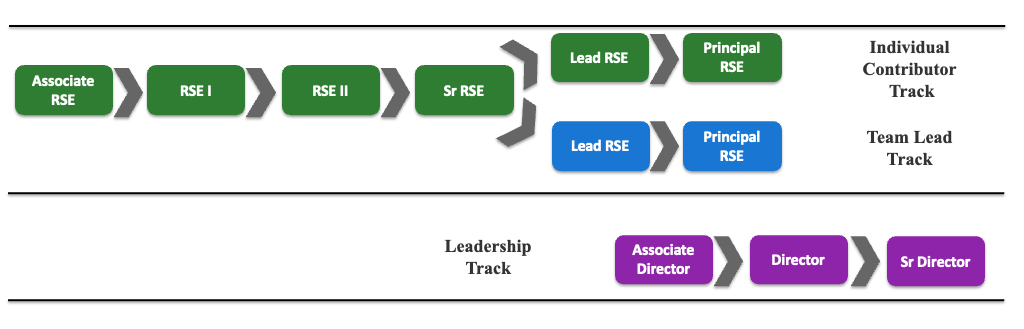}
    \caption{The Princeton RSE Career Ladder. The top (green) track represents the Individual Contributor (IC) Track. Positions along the IC track have no formal management or supervision responsibilities, the middle (blue) track splits from the IC track to include formal management. The bottom (purple) track represents the RSE group leadership positions and are located on an entirely separate track from the RSEs above.}
    \label{fig:ladder}
\end{figure*}

\section{Background and Motivation}
 \label{motivation}
Princeton’s RSE group operates centrally and aligns staff within research labs across disciplines with a model that has evolved over several years.
In 2016, the first “Research Software Engineer” role at Princeton University was created, just four years after the title took root at universities in the UK \cite{Baxter}.
In 2020, the Senior RSE position was established to recognize growth and provide a promotion path.
By 2022, however, rapid expansion of the RSE program, combined with early retention challenges and a limited path for promotion and growth for high performing RSEs prompted a comprehensive and intentional redesign.
These pressures were amplified by the growing demand for RSEs in the private sector that added to risk of turnover as talented RSEs were recruited to competitive industry positions.

In late 2022, at the time of the career ladder implementation, the group comprised 14 RSEs, two newly hired Associate Directors, and one Senior Director, with funding to expand by at least 10 new positions.
By 2026, the group has grown to 36 RSEs working in 23 academic units supported by a leadership team of three Associate Directors and one Senior Director.
The group is supported by two funding models and guided by a published partnership framework \cite{rse-partnership-guide}.
Details about the group’s operational model, including funding models, is described in \cite{princeton-rse-group}.
Strategic guidance for the funding and operational models is provided by the RSE Steering Committee, a group of appointed faculty and Research Computing leadership. 

Facing inconsistencies in job description language and pay scales, and struggling with the loss of RSEs to industry, Princeton partnered with an external consultant to standardize job descriptions, benchmark these roles against market standards, and map each role to a standard leveling framework \cite{radford-job-leveling} to align with HR structures.
Starting with the two existing RSE roles (RSE and Senior RSE) and the two existing leadership roles (Associate Director and Director), standard language was crafted to be domain-agnostic yet sufficiently detailed to preserve job level integrity to enable straightforward HR processes for position creation.
In completing this effort for the existing roles, it was acknowledged that the existing two-step career ladder was insufficiently able to support the predicted growth and efficacy of the team.
To resolve this disparity, a new career ladder \ref{career-ladder-architecture} was built by creating additional positions (plus team lead track variants) to enable earlier-career entry into the RSE path as well as continued recognition of advanced expertise and contribution to research, with or without people management.
The ultimate goal was to establish a complete career ladder that spanned entry level to expert while providing multiple meaningful career growth opportunities between them.

\section{Design Principles}
 \label{design-principles}
The group’s embedded partnership model required an RSE career ladder that could function across heterogeneous domains, align cleanly within existing overarching HR structures, and recognize scholarly impact beyond traditionally required publications.
It also needed to support both deep technical progression and leadership roles without forcing RSEs into people management as the only path for advancement.

The ladder was built on several core principles.
First, progression is based on both experience and competency, with expectations articulated across various key dimensions such as technical excellence, software engineering rigor, research impact, mentorship, leadership, and community engagement.
Language was standardized and graduated between levels, moving from “learn” and “contribute” at early levels to “define” and “innovate” at senior levels.

Second, the structured path offers a split pathway for advanced career RSEs after the level of Senior RSE.
A robust individual contributor track sets the foundation from Associate RSE through Senior RSE.
Lead and Principal RSEs can continue along an individual contributor track or follow a parallel team-lead and managerial track.
This split ensures that advancement does not require moving into people management and enables continued career growth maximizing an individual’s strengths and interests.

Third, alignment with HR was critical.
Defining a standardized RSE job description that aligns with role-specific research needs and enables successful and efficient hiring was a challenge.
It called for the understanding that requirements within each level can vary by position from aspects like domain knowledge to specific programming language or framework proficiency.
Furthermore, the software engineering industry is a particularly dynamic, rapidly evolving space where new technologies and methodologies regularly emerge and cause paradigm shifts (e.g., agile development, cloud computing, AI-assisted coding, etc.).
As such, standardized job descriptions must be adaptable to both the local environment (e.g., a particular research domain area) and the constantly shifting global software engineering industry.
Beyond the predictable technical job requirements, like domain knowledge or programming language fluency, we also strove to explicitly recognize the often “hidden” job requirements \cite{hidden-job-requirements} of research software engineering roles like creativity, communication skills, and the ability to navigate the ambiguous and rapidly-shifting academic research environment.
Writing a standardized job description that accommodated varied essential qualifications and skills while ensuring the skill range was properly compensated by market value took extensive collaboration and review. This approach challenged the model HR successfully uses with most staff positions where roles are uniquely defined with detailed requirements and the essential qualifications are always the same.
Princeton’s engagement with an external consulting company specializing in market comparisons for technical roles to review and support the standardization of job descriptions, identify current market value, and map each role to standardized levels supported consistency and alignment of RSE roles with HR standards.
This framework provides six levels of individual contributor: Entry (P1), Developing (P2), Career (P3), Advanced (P4), Expert (P5), and Principal (P6).
Given the responsibilities and technical complexity of the RSE role plus the education and experience requirements described in the standard job descriptions, the entire span of RSE roles mapped to the P3-P6 levels. 

Fourth, equity guided the process.
Consistent language and expectations are set within each role with clear paths up the ladder.
To accommodate the varied use of RSEs across all disciplines, there are placeholders for application of domain expertise, coding, and engineering proficiency that may vary between individual research labs and these placeholders are structured to promote fairness.
Accommodating established funding models and embedded partnerships requires clear alignment for fair management of RSEs across departments.
While RSEs are centrally managed, their work assignments come from PIs in their lab or department. Understanding the varied requirements within research labs and departments while also being able to clearly track growth of skills is critical for a fair promotion process. 

Finally, industry best practices and peer organizations also informed the design. We aimed to be consistent with NCSA’s research programmer career ladder \cite{rse-group-models-ncsa}, however this was  not always possible given the different priorities, expectations, and institutional frameworks.
The implementation was guided by the process outlined in The Manager’s Path \cite{managers-path-book}, emphasizing inclusive drafting with team feedback, benchmarking against external examples, aligning levels to market benchmarks consistent with internal HR salary bands, defining break-point levels, and maintaining separate technical and managerial tracks.
These principles reinforced Princeton’s approach and helped avoid common pitfalls.

\section {Career Ladder Architecture}
 \label{career-ladder-architecture}
The Princeton career ladder for RSE establishes a solid foundation for individual contributors, a branch for team leads (hybrid technical/people management roles), and an overarching management line (director roles).
To ensure consistency in expectations and in mapping duties to different levels across the ladder, we partition job responsibilities into the following categories:
\begin{enumerate}
\item \textbf{Application of Domain Expertise} - the degree to which a position needs to apply subject-matter knowledge to frame problems, define requirements, communicate with researchers, and validate results in order to ensure that software solutions/methods align with research priorities.
\item \textbf{Research Software Engineering} - the degree of necessary complexity of the software design while adhering to modern software engineering best practices. This can include architecture and design, version control, testing, CI/CD, documentation, packaging, performance tuning, etc. 
\item \textbf{Professional Development} - this reflects the program leadership’s explicit prioritization and encouragement of growth and development for RSEs in all relevant skillsets. Continuous learning is one characteristic that distinguishes great software engineers from the rest \cite{li-ko-begel}. Most Princeton RSE positions have some percentage of effort allocated to professional development (the more junior the position, the higher the percentage). Development can come through self-directed learning, formal training, and problem-based growth opportunities that arise organically during normal project work. 
\item \textbf{Technical Leadership} - the expectation of a role to provide direction, both within individual project contexts and with junior RSEs more broadly, on technical approaches and methodologies. This can include consulting on design, frameworks and approaches, code review, and/or teaching workshops.
\item \textbf{Management} - the formal responsibility to supervise employees which involves supporting RSEs in their technical work and facilitating communication with project partners. It can also include some project management and tracking. 
\item \textbf{Department Outreach and Collaboration} - the degree to which a role must build and maintain partnerships across departments, institutes, research teams, and individual Principal Investigators (PIs). Understanding partner needs is a crucial ingredient to a fruitful collaboration. 
\item \textbf{Strategic Vision} - the degree of responsibility a role has to focus on and develop long-term, forward-thinking strategy that anticipates and balances research and technology trends and institutional needs and priorities. 
\end{enumerate}

\begin{figure*}[t]
    \centering
    \includegraphics[width=0.8\textwidth]{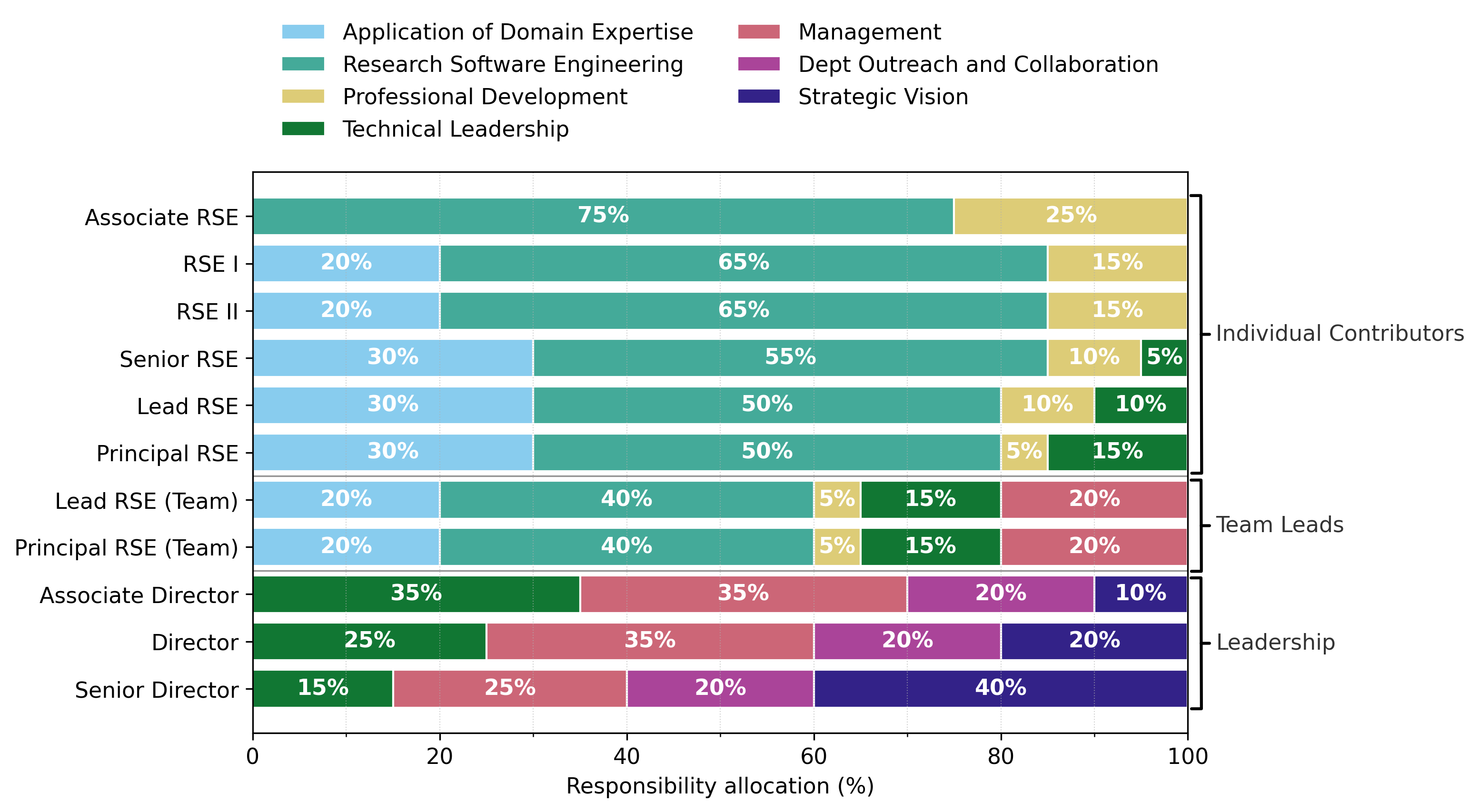}
    \caption{Responsibility distribution across all roles. These were the approximate percentages used in standard job descriptions and leveling.}
    \label{fig:staked-responsibility-plot}
\end{figure*}

The allocated percentages for each of the seven main responsibility categories across each standard level is shown in Figure \ref{fig:staked-responsibility-plot}. 
To further distinguish between levels, expectations of specific competencies essential to the roles were defined along the following three dimensions:

\begin{enumerate}
\item \textbf{Autonomy/Scope} - the effectiveness of the individual to independently define and execute work. It can range from completing well-scoped (by someone else) tasks to shaping project direction and guiding teams. 
\item \textbf{Task Complexity} - the complexity, size, ambiguity, and difficulty, of the problems and projects tasked to an individual. 
\item \textbf{Communication} - the skill and ability to communicate effectively across a broad range of stakeholders, teammates, and collaborators at appropriate levels of technical detail and domain specific language. Multiple modes of communication are considered including speaking, listening, writing, and visual.
\end{enumerate} 

The individual contributor track spans Associate, RSE I, RSE II, Senior, Lead, and Principal levels (Refer to the top/green track in Figure \ref{fig:ladder}).
Expectations progress from entry-level (highly guided RSE team-based work) to independent (execution guided by clear objectives), culminating in expert-level translation of research priorities into effective software solutions.
Years of experience and programming expertise increase correspondingly across all tiers.
Above the Associate level, education requirements and preferences are generally centered on computational disciplines with a preference for advanced degrees or equivalent research experience.

Team-lead variants (refer to the middle/blue track in Figure \ref{fig:ladder}) start at the Lead level with calculated decreases in responsibility for application of domain expertise, research software engineering, and professional development responsibilities to accommodate added responsibilities for technical leadership (scope-setting, milestone ownership, and mentoring of earlier-career RSEs).
Lead and Principal RSEs can officially manage up to three RSEs. 

Associate Director, Director, and Senior Director are the three administrative leadership roles.
These roles prioritize technical leadership, people management, outreach and collaboration, and the strategic vision for the RSE group.
We have included these roles to highlight the management tiers needed to properly support a growing RSE group of Princeton’s size, but de-emphasized discussion given their strong dependence on institutional criteria rather than technical role specification. 

Responsibility distribution (Figure \ref{fig:staked-responsibility-plot}) varies by level. Early roles (Associate, RSE I, and RSE II) devote most effort to application of domain expertise, research software engineering, and professional development while senior positions (Senior RSE, Lead, Principal) take on more technical leadership and mentoring of earlier-career RSEs.
Team-lead roles (Lead, Principal) introduce limited people management responsibilities, while directors (Associate Director, Director, Senior Director) handle substantial people management responsibilities, as well as organizational strategy and outreach.

\section{Language Framework}
 \label{language-framework}

Building a comprehensive ladder for technical positions that not only covered the intricate needs of the RSE role, but also clearly defined the path to promotion, required intentional consideration and structure around the language used in job descriptions.
Inconsistent descriptors create ambiguity, inequity, or simply confusion.
This was addressed by standardizing responsibility categories across all roles and then adopting a grid of graduated verbs and adjectives representing key essential responsibilities.
Significant discussion and debate was had over specific words to ensure proper representation of what was needed to qualify for promotion and the appropriate placement of existing RSEs in the previous structure.
This approach was time consuming, but aligned with clear HR standards for market value assessment and made expectations legible and future promotion discussions fairer, reducing misinterpretation and confusion.

\begin{figure*}[t]
    \centering
    \includegraphics[width=1.0\textwidth]{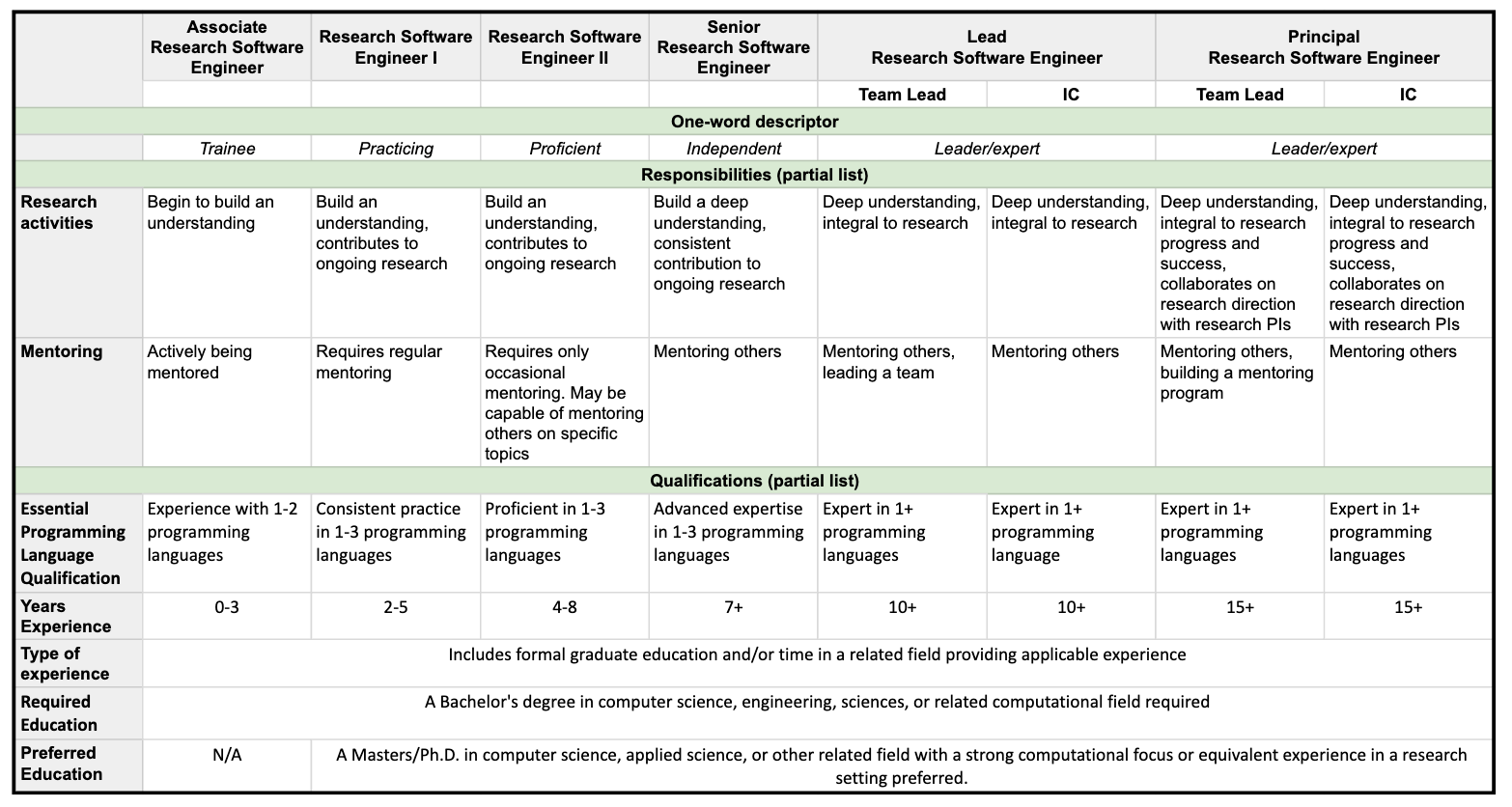}
    \caption{Example responsibilities and qualifications for RSE career levels. Notes this is a subset of only two example responsibilities and qualifications. This demonstrates the relative changes between levels, rather than an exhaustive description. Lead and Principal levels are split into the Team Lead, with people management responsibilities, and Individual Contributor (IC) tracks.}
    \label{fig:responsibilities-matrix}
\end{figure*}

Figure \ref{fig:responsibilities-matrix} highlights two representative responsibilities across all the levels: mentoring and research activities.
The grid shares the clear expectations of giving/receiving mentorship to advance in the RSE role.
From `actively being mentored' to `mentoring others and building a mentoring program', there are incremental advances as each role is defined up the career path.
Similarly, the expectation around the participation in research activities grows along the ladder with language to match.

\section{Implementation Process}
 \label{implementation-process}
The implementation unfolded in four phases: Discovery, Drafting, Stakeholder Engagement and Calibration, and Rollout. 
During discovery, we focused on collecting and analyzing existing RSE job descriptions, identifying pain points in hiring and promotion, and gathering feedback from RSEs, faculty, leadership, and HR partners.
Leveraging what was learned in discovery, we started drafting a desired future state for the RSE group.
This was supported by creating competency matrices and domain-agnostic job descriptions. During this phase, Princeton engaged an external consulting agency for benchmarking and mapping to standardized job levels.
Feedback from the continued collaboration with the consulting agency was incorporated in the iterative drafts as the framework became more clear.  

Once the drafts were stable, stakeholder engagement and calibration followed.
This required HR reviews and approval, pilot evaluations, proposed calibration of existing RSEs from the previous levels to the newly created levels, and faculty input, culminating in approval from University leadership and the RSE Steering Committee.
Throughout this phase, stakeholder engagement was critical.
This included engagement with multiple levels of administrative departments from HR to ODFR to the Office of Information Technology (OIT), as well as with key faculty stakeholders already involved with the RSE program, and ultimately the RSE Steering Committee.
This effort paralleled an expansion of funding for the RSE program so it was imperative to present the proposed strategy and get full buy-in to ensure clarity before the new positions were created, posted, and hiring began.
The revised RSE career path and updated salary ranges set by market value did impact incumbent RSEs.  Confirming all adjustments to the existing model for funding and hiring RSEs was an integral step in receiving final approval to roll out this new model.
Funding support from the University and changes in the central cost sharing model were necessary in some cases to ease the transition to new salary ranges, especially for RSEs supported by grant funding.  

Following approval of the new RSE career path, the rollout began.
Officially mapping incumbents to new titles, creating detailed communication plans for partnering departments, PIs, and RSEs, and strategizing specifically timed conversations with impacted parties to ensure clarity during the transition was vital.
RSE hiring managers were then empowered and authorized to create new RSE positions using the standard job descriptions, matching to unique position need and fit.
Now new RSE positions are straightforward to create and route for approval. Starting with the standard job description at any given RSE level, hiring managers can  adapt (typically less than 20\% of) the language to include domain specifics and other technical details inherent to each unique RSE position.
Iteration continues annually to refine descriptors, address unintended consequences, and develop a formal promotion guide.

\section{Discussion}
 \label{discussion}

\begin{figure*}[t]
    \centering
    \includegraphics[width=0.8\textwidth]{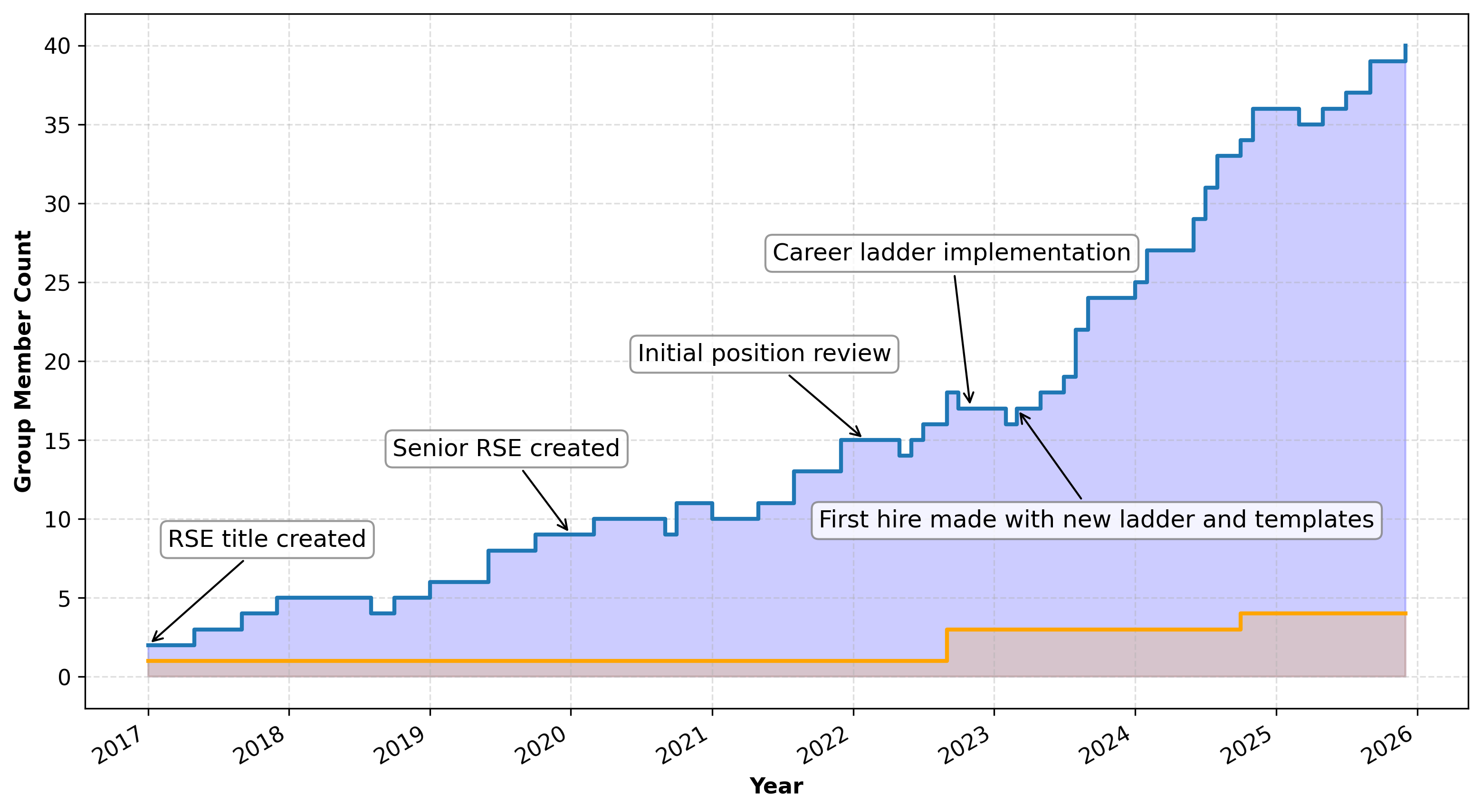}
    \caption{The Princeton RSE Group membership over time annotated with career ladder milestones. The blue line represents a count of all individuals in the RSE Group, including RSEs of all levels and the leadership team of directors. The orange line represents a count of the subset in the leadership track. Upticks represent new hires and additions while downticks represent departures.}
    \label{fig:group-member-count}
\end{figure*}

Since its adoption, the new career ladder has received positive reception among RSEs, streamlined HR processes for position creation, and benefited recruitment and retention.
To compare pre- and post-rollout retention effectiveness we calculate an average turnover percentage as the number of RSEs who left the group divided by the average number of group members for that time period.
In the five years preceding the November 2022 career ladder rollout, six RSE Group members left Princeton University to outside competitors. With an average group size of 8.7 RSEs during that time, this corresponds to a 70\% turnover rate for the pre-rollout group.
In the three years since the rollout, two RSE Group members have left to outside competitors. With an average group size of 28.8 members, this corresponds to a roughly 7\% turnover.
This massive reduction in turnover is unlikely to be solely attributed to the adoption of the new career ladder, and is offset by the team's rapid growth over this time period.
However, alignment of the RSE roles with market value and the positive reception with RSE group members gives us confidence that the new career ladder has been a substantial factor in the improved retention.

We polled members of the RSE Group in 2021 (before) and in 2025 (after) the adoption of the career ladder.
 The 2021 feedback indicated that the lack of a clear career path was a top area for improvement.
Following the implementation of the 2022 career ladder and allowing time for the impacts to be understood, the feedback from  2025 group members clearly showed that the RSE career structure and professional development were positively received.
The 2025 responses also demonstrate that there remains opportunity for continued improvement and clarity around promotion paths.
For example, some RSE positions are tied to one or more grant-funded projects and those projects may not need (or have the financial flexibility to support) RSEs at more advanced levels on the career ladder.
This reinforces the need for a formal promotion guide that takes into consideration potential limitations to career growth and underscores that the ladder will continue to evolve over time.

Improved hiring efficiency, especially around new position creation and related approval processes, was a crucial and immediate benefit of this implementation.
After the creation of standard templates and RSE levels, approval and grading processes became streamlined within the University, thanks to the new shared vocabulary and understanding of the RSE role within the University context. Approval processes that previously could take multiple months, due to the varied and complex technical nature of RSE positions, now took weeks. With the newly funded roles and rapid growth, this benefit enabled Associate Directors to hire at a much more aggressive rate than would have been previously possible. 

Princeton’s experience highlights several trade-offs.
Competency frameworks must standardize role responsibilities and expectations but also accommodate diverse domains while simultaneously providing clear measures to satisfy HR leveling frameworks.
Technical mastery should be valued on par with managerial roles to avoid perceived hierarchy removing requirements for people management to advance.
An RSE’s outputs (e.g., software packages, feature additions, bug fixes, publications, etc.) and the scale of these outputs can vary significantly by project, requiring careful consideration for performance evaluation and promotion between ranks.
Projects can also differ significantly with respect to technical and personnel requirements and funding resources, which can substantially impact the career path of an RSE attached to said project.
For example, while our career ladder offers a track for individual contributions and a parallel track for team leads, not every project has the capacity for multiple RSEs, so an RSE seeking growth into the team lead track may have to switch projects.
Role changes and natural responsibility drift remains a persistent challenge, necessitating regular reviews to maintain clarity, fairness, and the incorporation of new trends and requirements. 

Finally, because the change would impact administrative and academic departments and research labs across all disciplines, understanding Princeton’s culture and institutional norms was essential to shaping a solution and implementation approach that would be effective across the University.
By working with such a broad group of partners and stakeholders to implement the RSE career ladder redesign in a centralized, University‑accepted manner, the path was also established for departments not tied to  the central RSE Group to adopt the new titles and career ladder with ease.

\section{Adoption Recommendations}
 \label{adoption-recommendations}
Princeton is a relatively small, centralized research university that emphasizes undergraduate and doctoral education. There are no professional schools such as medicine, law, or business.
This structural context shapes how institution‑wide initiatives can be designed and implemented. 
What was successful for Princeton may not be successful at larger, decentralized institutions.
It is vital to understand the institutional culture and involve the correct people from faculty, administration, and leadership in the discovery and design process from the beginning. Institutions seeking to replicate this model should begin with an audit of their existing roles and pain points, documenting a clear ‘as-is’ starting point.
Following a comprehensive audit, draft the vision and goals for this effort to clarify the desired ‘to-be’ future state. Understanding the gap between the starting point and desired end point prior to defining the implementation steps will provide a solid outline of the effort needed and help identify critical stakeholders to be included. 
To support this, draft domain-agnostic, competency-based descriptions using graduated language and align those with HR job families and market benchmarks, which is essential for sustainability.
Engage with RSE and campus partners to ensure vision and strategy are met with acceptance and support to prevent future resistance.
Clear planning and communication during rollout, especially when existing staff and PIs are impacted, is crucial to implementation success.
Finally, to refine and sustain the RSE career ladder over time, allocate time to annually iterate on the career ladder, factoring in data like RSE role retention, market trends, and progression metrics.

\section{Conclusion}
 \label{conclusion}
The RSE career ladder implementation case study highlights Princeton's understanding of how indispensable Research Software Engineers are to computational research and scholarship. The University's investment in and commitment to designing a meaningful career path to recruit, develop, and retain RSE talent is a testament to the value RSEs bring to the modern academic research enterprise. Strategically defining a progressive career path that standardized job descriptions to align with existing HR frameworks while maintaining flexibility to recognize the unique nature of individual RSE positions took extensive effort and collaboration with numerous academic and administrative partners. The end product was a transparent, competency-based, RSE career ladder mapped to market value that materially improved hiring efficiency, retention, and added clarity to the pathway for professional growth. These improvements are evident through the successful growth and maintenance of Princeton's RSE group to 36 RSEs across 23 academic units. A career ladder spanning Associate through Principal levels with a parallel team-lead track provides a strong roadmap for continued progression in technical expertise and contribution to research without requiring people management. Princeton University's RSE stakeholders will continue to reflect on the RSE career ladder, learn from peers with similar and different RSE models, and proactively iterate on the model to support career growth of Research Software Engineers. We expect that this intentionally designed and robust career ladder will provide RSEs with long-term stability and career growth opportunities, further enabling the advancement of computational research and scholarship at Princeton. 

\section{Acknowledgments}
 \label{acknowledgment}
We extend our sincere gratitude to the many Princeton colleagues and partners who contributed to the development of the RSE career ladder. A special thank you to the RSE Steering Committee for your continued guidance and support during this work.
The effort benefited from the insight, time, and thoughtful engagement of collaborators across Human Resources, Office of Information Technology, Princeton Institute of Computational Science and Engineering (PICSciE), the Office of the Dean for Research and academic units across campus.
We are also profoundly grateful to all of the members of the Princeton RSE group, past and present. Their sustained excellence, creativity, and commitment to a diverse set of research projects provided both the motivation and foundation for the career ladder.

We also gratefully acknowledge and thank Curt Hillegas, Jeroen Tromp, Jay Dominick, Pablo Debenedetti, and Peter Schiffer for their longstanding support of the RSE program at Princeton. 
Their advocacy, leadership, and belief in the importance of the RSE mission have made this work possible and continue to shape the environment in which RSEs can thrive and make lasting research impact.

\bibliographystyle{elsarticle-num}
\bibliography{references}

\end{document}